\documentclass[conference, rgb]{IEEEtran}
\IEEEoverridecommandlockouts

\usepackage{cite}
\usepackage{amsmath,amssymb,amsthm}
\usepackage{algorithmic}
\usepackage{graphicx}
\usepackage{textcomp}
\usepackage{xcolor}
\usepackage{tikz}
\usepackage{pgfplots}
\pgfplotsset{compat=1.17}
\usepackage{hyperref}

\newcommand{\SU}{\operatorname{SU}}

\newcommand{\SWAP}{\operatorname{SWAP}}

\begin{document}

\title{Clique detection using symmetry-restricted quantum circuits
}

\author{\IEEEauthorblockN{Maximilian Balthasar Mansky\IEEEauthorrefmark{1}\IEEEauthorrefmark{2}, Tobias Rohe\IEEEauthorrefmark{1}, Dmytro Bondarenko\IEEEauthorrefmark{1},\\ Linus Menzel\IEEEauthorrefmark{1} and Claudia Linnhoff-Popien\IEEEauthorrefmark{1}}
\IEEEauthorblockA{\IEEEauthorrefmark{1}Institute of Informatics, LMU Munich}
\IEEEauthorblockA{\IEEEauthorrefmark{2}
Email: maximilian-balthasar.mansky@ifi.lmu.de}}

\maketitle

\begin{abstract}
We show the application of permutation-invariant quantum circuits to the clique problem. The experiment asks to label a clique through identification of the nodes in a larger subgraph. The permutation-invariant quantum circuit outperforms a cyclic-invariant alternative as well as a standard quantum machine learning ansatz. We explain the behavior through the intrinsic symmetry of the problem, in the sense that the problem is symmetric under permutation of both the feature and the label. 
\end{abstract}


\section{Introduction}

Quantum computing has emerged as a promising computational paradigm, attracting considerable attention in recent years. With demonstrated advantages over classical algorithms \cite{shorAlgorithmsQuantumComputation1994, groverFastQuantumMechanical1996, deutsch_rapid_1992}, there is a growing expectation that quantum computing will eventually surpass classical methods. However, recent developments have seen a slowdown in new quantum algorithms, with research momentum shifting toward the domain of quantum machine learning \cite{schuldIntroductionQuantumMachine2015, biamonteQuantumMachineLearning2017}. In this framework, quantum circuits are built using parameterized gates, which can be tuned to approximate target functions.

Unlike classical machine learning, quantum machine learning leverages a fundamentally different set of operations for function approximation. At a high level, classical models are typically composed of linear transformations interleaved with non-linear activation functions, structured into layers. In contrast, quantum circuits impose different constraints on the learning model. Parameters are usually embedded into single-qubit gates, while entanglement is introduced via non-parameterized two-qubit gates that connect qubits. Instead of nested layers, quantum circuits follow a multiplicative composition, shaped by the mathematical formalism of quantum mechanics \cite{nielsenQuantumComputationQuantum2010, hallLieGroupsLie2013}.

This structural distinction allows for the design of specialized quantum machine learning circuits, often referred to as ansatzes, tailored to specific tasks. In classical machine learning, architectural tuning or incorporation of specific layers can adapt the model to a problem’s structure. For instance, convolutional neural networks \cite{LeCunbackpropagationAppliedHandwritten1989} are explicitly constructed to encode translational invariance. More recent innovations, such as attention mechanisms \cite{vaswani_attention_2017}, provide models with the capacity to learn relevant symmetries from data.

Quantum machine learning has not yet reached this level of flexibility, and incorporating meaningful symmetries into quantum models remains an open challenge \cite{mansky_permutation-invariant_2023, laroccaGroupInvariantQuantumMachine2022, meyerExploitingSymmetryVariational2023}. Owing to their unique structure, quantum circuits may enable learning paradigms that go beyond classical strategies. Classical networks express symmetry primarily through their overall architecture—as seen in convolutional networks \cite{dhillon_convolutional_2020}, graph neural networks \cite{gori_new_2005}, and, to some extent, transformer-based models \cite{vaswani_attention_2017, bronsteinGeometricDeepLearning2021}—but typically do not encode specific symmetries of individual problems.

Quantum circuits, on the other hand, allow for the exact encoding of problem-specific symmetries through their multiplicative structure. If the symmetry underlying a physical system or computational problem is well understood—either from empirical insights or theoretical analysis—it can be directly embedded into the circuit’s design. This approach enables a more precise alignment between the quantum circuit ansatz and the problem’s inherent structure, which can in turn enhance model performance.

In this work, we examine the application of symmetrized quantum circuits to the Max-clique problem from computer science. A clique within the graph is a set of nodes that forms a complete graph – each node is connected to each other node in the set.\cite{grossHandbookGraphTheory2004}. We utilize a standard graph embedding \cite{mansky_solving_2025} and ask the quantum circuit to label all nodes that are part of the subgraph.

\section{Related work}

In the area of classical neural networks, graph problems are generally solved with graph neural networks \cite{scarselli_graph_2009}, which operate on the individual edges of a graph. This has found significant application across graph problems, able to approximately solve many hard problems on graphs \cite{zhouGraphNeuralNetworks2020, yangHyperbolicGraphNeural2022}. There is already a quantum analogue of this classical approach, the quantum graph neural network \cite{verdon_quantum_2019}, with similar applications to graph problems \cite{ceschini_graphs_2024}. Both approaches work on the individual edges. In this model, an isomorphic graph is a different graph and can give different results.

In contrast, there exists the approach of integrating the global symmetry of the problem into the system. This is termed group-invariant (sometimes -equivariant) quantum computation \cite{meyerExploitingSymmetryVariational2023, ragoneUnifiedTheoryBarren2024}. This leads to interesting mathematics \cite{mansky_scaling_2024, schatzkiTheoreticalGuaranteesPermutationequivariant2024, nguyenTheoryEquivariantQuantum2024} and to its successful application to graph problems \cite{mansky_solving_2025}. The structural difference here is important – graph neural network forego a larger symmetry for solutions on particular graphs, while group-invariant approaches incorporate the larger symmetry of isometries.

\section{Quantum circuits}

The quantum circuits used in the experiments reflect discrete symmetries between the qubits. Discrete symmetries can always be expressed through permutations. We realize the permutations on the quantum circuits through $\SWAP$ operations between qubits \cite{mansky_permutation-invariant_2023, mansky_scaling_2024}. A quantum circuit can be made invariant under the symmetry \cite{mansky_permutation-invariant_2023, mansky_solving_2025} In brief, the restriction on the symmetry can be transported to the Lie algebra, the generator space. Here, generators for quantum circuit elements can be grouped together to fulfill the symmetry requirements. The generators can be transported back to the quantum circuits to create permutation-invariant quantum circuits. \cite{laroccaGroupInvariantQuantumMachine2022, ragoneUnifiedTheoryBarren2024,  schatzkiTheoreticalGuaranteesPermutationequivariant2024} The quantum circuits are shown in table \ref{tab:circuits}

\begin{table}
\caption[Quantum circuit building blocks]{The quantum circuits building blocks used in the quantum machine learning analysis. Each quantum circuit diagram shows one layer of the quantum circuit. This layer is repeated until the desired number of parameters is achieved. Gates with the same color contain shared parameters. $U$ gates indicate gates  that cover the whole $\SU(2)$ sphere. From top to bottom: Permutation-invariant, cyclic invariant and strongly-entangling layer.}\label{tab:circuits}
\centering

\begin{tikzpicture}[gate/.style={rectangle, draw=black, fill=white, inner sep=3pt}, xscale=.53, yscale=.45, baseline=(current bounding box.center)]
\path (2,1) -- (2, 7);
\foreach \i in {1,2,...,6} {
\draw (0.5, \i) --+ (17, 0);
\draw (1, \i) node[gate, fill=red!20] {X}
	(2, \i) node[gate, fill=orange!20] {Y};
	\foreach \j in {3, 4, ..., 8} {
	\node (p\j\i) at (\j, \i) {\phantom{$ZZ$}};};};
\foreach \i in {1,2, ..., 5} {
\draw[line width=.3pt, line cap=rect, double=purple!20, double distance=12pt] (\i + 2, 1) --+ (0, \i);
\draw (\i + 2, 1) node {ZZ};};

\foreach \i in {1,2, ..., 4} {
\draw[line width=.3pt, line cap=rect, double=purple!20, double distance=12pt] (\i + 7, 2) --+ (0, \i);
\draw (\i + 7, 2) node {ZZ};
\draw (\i + 2.5, \i + 1) --+ (5 - \i, 0);};

\foreach \i in {1,2, ..., 3} {
\draw[line width=.3pt, line cap=rect, double=purple!20, double distance=12pt] (\i + 11, 3) --+ (0, \i);
\draw (\i + 11, 3) node {ZZ};
\draw (\i + 7.5, \i + 2) --+ (4 - \i, 0);};

\foreach \i in {1,2} {
\draw[line width=.3pt, line cap=rect, double=purple!20, double distance=12pt] (\i + 14, 4) --+ (0, \i);
\draw (\i + 14, 4) node {ZZ};
\draw (\i + 11.5, \i + 3) --+ (3 - \i, 0);};

\foreach \i in {1} {
\draw[line width=.3pt, line cap=rect, double=purple!20, double distance=12pt] (\i + 16, 5) --+ (0, \i);
\draw (\i + 16, 5) node {ZZ};
\draw (\i + 14.5, \i + 4) --+ (2 - \i, 0);};
\end{tikzpicture}

\begin{tikzpicture}[gate/.style={rectangle, draw=black, fill=white, inner sep=3pt}, xscale=.6, yscale=.45, baseline=(current bounding box.center)]
\path (2,1) -- (2, 7);
\foreach \i in {1,2,...,6} {
\draw (0.5, \i) --+ (14.0, 0);
\draw (1, \i) node[gate, fill=red!20] {X}
	(2, \i) node[gate, fill=orange!20] {Y};
	};
\foreach \i in {1,2, ..., 5} {
\draw[line width=.3pt, line cap=rect, double=purple!20, double distance=14pt] (\i + 2, \i) --+ (0, 1);
\draw (\i + 2, \i) node {ZZ};};
\draw[line width=.3pt, line cap=rect, double=purple!20, double distance=14pt] (8, 1) --+ (0, 5);
\draw (8,1) node {ZZ};
\foreach \i in {2, 3, ..., 5} {
\draw (7.5, \i) --+(1,0);};
\foreach \i in {1,2, ..., 4} {
\draw[line width=.3pt, line cap=rect, double=violet!20, double distance=14pt] (\i + 8.0, \i) --+ (0, 2);
\draw (\i + 7.5, \i+1) --+ (1,0);
\draw (\i + 8.0, \i) node {ZZ};};
\foreach \i in {1,2} {
\draw[line width=.3pt, line cap=rect, double=violet!20, double distance=14pt] (12.0 + \i, \i) --+ (0, 4);
	\foreach \j in {1, 2, 3} {
	\draw (11.5 + \i, \i + \j) --+ (1,0);};
\draw (12.0 + \i,\i) node {ZZ};};
\end{tikzpicture}

\begin{tikzpicture}[gate/.style={rectangle, draw=black, fill=white, inner sep=3pt}, xscale=.6, yscale=.45, baseline=(current bounding box.center)]
\path (2,1) -- (2, 7);
\begin{scope}[yscale=-1, yshift=-7cm]
\foreach \i/\col in {1/0, 2/20, 3/40, 4/60, 5/80, 6/100} {
\draw (0.5, \i) --+ (14, 0);
\draw (1, \i) node[gate, fill=red!\col!violet!20] {U};
	};
\foreach \i in {5,4, ..., 1} {
\path (\i + 1, \i+1) node[circle, draw=black] (target) {} (\i + 1, \i) node[circle, fill=black, inner sep=1.3pt] (control) {};
\draw (target.south) -- (control.center);};
\path (7, 1) node[circle, draw=black] (target) {} (7, 6) node[circle, fill=black, inner sep=1.3pt] (control) {};
\draw(target.north) -- (control.center);

\begin{scope}[xshift = 7cm]
\foreach \i/\col in {1/0, 2/20, 3/40, 4/60, 5/80, 6/100} {
\draw (1, \i) node[gate, fill=orange!\col!purple!20] {U};
	};
\foreach \i in {1,2, ..., 4} {
\path (\i + 1, \i+2) node[circle, draw=black] (target) {} (\i + 1, \i) node[circle, fill=black, inner sep=1.3pt] (control) {};
\draw (target.south) -- (control.center);};
\foreach \i in {1,2} {
\path (5 + \i, \i) node[circle, draw=black] (target) {} (5 + \i, 4 + \i) node[circle, fill=black, inner sep=1.3pt] (control) {};
\draw(target.north) -- (control.center);};
\end{scope}
\end{scope}

\end{tikzpicture}

\end{table}

We use different symmetries in our quantum circuits. The strongest symmetry available on a qubit level is the permutation symmetry $S_n$ of all possible permutations $\pi_k$. Each two-qubit gate needs to connect from each qubit to any other qubit in this  setting. The next weaker regular symmetry is the cyclic symmetry $C_n$. This symmetry is characterized by cyclically shifting each element to the next one, such that $1\to 2, 2\to 3$ and so on. This provides a baseline for the architecture. Lastly, we compare against one of the standard circuits of quantum machine learning, the strongly entangling layer implemented by \texttt{pennylane} \cite{bergholmPennyLaneAutomaticDifferentiation2022}. 

The different layer structures are concatenated until approximately the same number of parameters is reached. The target is 120 parameters, meaning 40 repetitions in the case of the permutation-invariant quantum circuits, 30 in the case of the cyclic-invariant quantum circuits and three for the strongly-entangling layer in the standard ansatz.

\section{Experiments}

We test our quantum circuits on the clique problem. Within a graph $\mathcal{G}$, a clique is a set of nodes $\{N\}$ that is fully connected, meaning that each node $N_i$ within the set is connected to every other node $N_j$ in the same set. The features of the data set are Erdős-Rényi random graphs \cite{erdos_random_1959} and the labels indicate the clique position. The graphs are sampled with a random edge connectivity and are then separated into the label groups. Graphs without a clique of at least the desired size are labeled as a blanket $-1$ expected measurement on all qubits. For graphs containing larger cliques, the contained clique is chosen randomly. 

This is mapped to the quantum computer with a graph embedding \cite{mansky_solving_2025} where each node in the graph is identified with a qubit and each edge is represented by a CZ gate. The label is a measurement pattern, such that a qubit is expected to be $1$ if it is part of the clique and $-1$ if it is not. The experiments are set up to detect a 4-clique in the case of 6 qubits and a 5-clique in the case of 8 qubits. Other clique sizes show similar results.

The quantum circuit essentially needs to learn to identify the fully connected subcircuit hidden in the entanglement operation of the CZ gates and then put out the corresponding amplitude-encoded labels for the measurement. With the different quantum circuit symmetries, it is possible to figure out whether the desired solution has a given symmetry, in which case convergence is possible, or whether it is not subject to that symmetry, in which case no convergence should be observed \cite{mansky_solving_2025}.  

\section{Results}

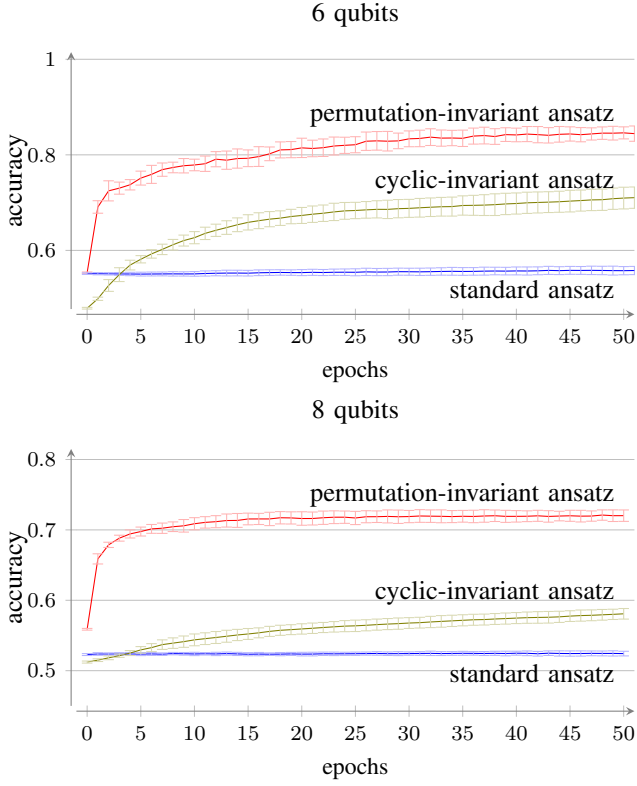
\begin{figure}
\caption[Validation set performance of quantum machine learning on select graph problems.]{The validation set performance of different quantum circuits on 6 and 8 qubits. The high performance of the permutation-invariant quantum circuit is clearly visible.}\label{tab:results}
\begin{tikzpicture}[baseline=(current bounding box.center)]
\begin{axis}[small,
height=5cm, width=9cm,
no markers,
xlabel = {epochs},
axis x line = bottom,
axis y line = left,
ymajorgrids,
major grid style = {very thin, gray!50},
major tick style = {very thin, gray!50},
axis line style={gray},
axis line shift=2pt,
xmin = -1,
xmax = 51,
ymax=1.015,
ylabel = {accuracy},
every axis y label/.style={at={(ticklabel cs:.5)},rotate = 90, anchor=center},
title = {6 qubits},
	]
	\addplot [red, error bars/.cd, y dir = both, y explicit, error bar style={ thin, red!30}] table [x=Epoch, y=Node_Avg, col sep=comma, y error = Node_Avg_Error]{data/Accuracy_Sn_Average_6_qubits.csv};
	\addplot [blue, error bars/.cd, y dir = both, y explicit, error bar style={ thin, blue!30}] table [x=Epoch, y=Node_Avg, col sep=comma, y error = Node_Avg_Error]{data/Accuracy_Entanglement_Average_6_qubits.csv};
	\addplot [olive, error bars/.cd, y dir = both, y explicit, error bar style={ thin, olive!30}] table [x=Epoch, y=Node_Avg, col sep=comma, y error = Node_Avg_Error]{data/Accuracy_Cn_Average_6_qubits.csv};

	\draw[red] (yticklabel cs: 0) -- (yticklabel cs: 1);
	\draw (axis cs: 50, .84) node[above left] {permutation-invariant ansatz}
		(axis cs: 50, .70) node[above left] {cyclic-invariant ansatz}
		(axis cs: 50, .48) node[above left] {standard ansatz};

\end{axis}
\end{tikzpicture}

\begin{tikzpicture}[baseline=(current bounding box.center)]
\begin{axis}[small,
height=5cm, width=9cm,
no markers,
xlabel = {epochs},
axis x line = bottom,
axis y line = left,
ymajorgrids,
major grid style = {very thin, gray!50},
major tick style = {very thin, gray!50},
axis line style={gray},
axis line shift=2pt,
xmin = -1,
xmax = 51,
ymin = .45,
ymax=.815,
ylabel = {accuracy},
every axis y label/.style={at={(ticklabel cs:.5)},rotate = 90, anchor=center},
title = {8 qubits},
	]
	\addplot [red, error bars/.cd, y dir = both, y explicit, error bar style={ thin, red!30}] table [x=Epoch, y=Node_Avg, col sep=comma, y error = Node_Avg_Error]{data/Accuracy_Sn_Average_8_qubits.csv};
	\addplot [blue, error bars/.cd, y dir = both, y explicit, error bar style={ thin, blue!30}] table [x=Epoch, y=Node_Avg, col sep=comma, y error = Node_Avg_Error]{data/Accuracy_Entanglement_Average_8_qubits.csv};
	\addplot [olive, error bars/.cd, y dir = both, y explicit, error bar style={ thin, olive!30}] table [x=Epoch, y=Node_Avg, col sep=comma, y error = Node_Avg_Error]{data/Accuracy_Cn_Average_8_qubits.csv};

	\draw[red] (yticklabel cs: 0) -- (yticklabel cs: 1);
	\draw (axis cs: 50, .72) node[above left] {permutation-invariant ansatz}
		(axis cs: 50, .58) node[above left] {cyclic-invariant ansatz}
		(axis cs: 50, .47) node[above left] {standard ansatz};

\end{axis}
\end{tikzpicture}

\end{figure}

For each quantum machine learning structure, we create a balanced dataset of 3000 graphs, split into a training and a test set. The training set contains only 100 graphs per epoch, with the remaining 2900 graphs forming the validation set. We use the quantum natural gradient \cite{stokes_quantum_2020} on the classification task. 

The results are presented in figure \ref{tab:results}. In both cases, the permutation-invariant quantum circuit performs well on the graph problem, with a fast convergence within a few epochs. The cyclic invariant quantum circuit also converges, albeit at a slower pace. In separate experiments, we find that the cyclic invariant quantum circuit converges to approximately 65\% accuracy after 200 epochs. The standard ansatz shows no signs of convergence, giving approximately random results throughout the training period. The error bars indicate the 95\% confidence interval averaged over 10 runs of simulations with random seeds.

\section{Discussion}

The experiments shown here have grown out of a larger study of the performance of symmetrized quantum circuits on a variety of settings, typically on problems that exhibit a symmetry. The quantum circuit is then matched to the symmetry \cite{mansky_permutation-invariant_2023, mansky_solving_2025}. This entails a graph problem formulated as a binary decision problem, such as asking whether a graph is connected or bipartite. In the present case, the problem is prima facie not invariant under a symmetry – the labeling of the nodes directly decides where the clique is located. This should lead to worse performance for permutation-invariant quantum circuits, to the point that they should not converge at all.

However, this overlooks the fact the presented problem is subject to a permutation symmetry, since both the feature, the incoming graph, as well as the label, denoting the clique, are permutation-invariant. In the case of a binary decision problem on a graph \cite{mansky_solving_2025}, the feature individually is permutation-invariant. Here, the permutation transformation has to apply to both the feature and the label,
\begin{equation}
    \mathcal{F} \to \ell \Leftrightarrow \pi_k\mathcal{F} \to \pi_k\ell
\end{equation}
for features $\mathcal{F}$, labels $\ell$ and permutations $\pi_k$. This realization opens the door to a deeper study of the phenomenon and broadens the applicability of permutation-invariant quantum circuits to cases where the symmetry is intrinsic, rather than extrinsic.

\section*{Acknowledgements}

MBM acknowledges funding from the German Federal Ministry of Education and Research (BMBF) under the funding program ”Förderprogramm Quantentechnologien – von den Grundlagen zum Markt” (funding program quantum technologies – from basic research to market), project BAIQO, 13N16089.

\bibliographystyle{IEEEtranS}

\end{document}